\documentclass{article}
\usepackage{frascatiphys_R}
\usepackage{fancybox}  
\usepackage{latexsym}
\usepackage{amsmath,amssymb}
\usepackage{latexsym,graphicx,a4,epsfig,psfrag,here}

\newcommand{\bea}{\begin{eqnarray}}
\newcommand{\eea}{\end{eqnarray}}
\newcommand{\fs}{\,\, .}

\newcommand{\scs}{\, , \,}
\newcommand{\bc}{\begin{center}}
\newcommand{\ec}{\end{center}}
\begin{document}
\title{ {\bf KAONIC ATOMS IN QCD}}
\author{J. Gasser\\
\em {Institute for  Theoretical Physics, University of Bern,}\\{\em CH-3012 Bern, Switzerland
}}
\maketitle
\baselineskip=11.6pt
\begin{abstract}
In this talk, I comment on the theoretical and experimental status of kaonic atoms, in particular
 $\bar K \pi$ and $\bar K p$ bound states. 
\end{abstract}

\baselineskip=14pt
\section{Introduction}
Kaonic  atoms are particular examples of {\em hadronic atoms}. They are of the type
$\bar K X$, with 
$X=\pi, K;  \text{p; d; $^3$He; $^4$He} \ldots$.
Kaonic atoms are by definition bound by electromagnetic interactions, so 
a more precise  title of my talk would be 
 {\em Kaonic atoms in QCD + QED}. On the other hand, 
 { deeply bound
  kaonic nuclear states} are of a different variety - as far as I understand, they are 
predicted to exist already in the framework
of QCD\cite{akaishi,outa}, electromagnetic forces are not required for
their formation. 
 I do not  consider these systems  here (nor  $\bar K K$ bound
states\cite{krewald}). The reason to investigate {\em hadronic atoms} in
general is 
the following: as just said, they are formed  by electromagnetic forces, which are well
known. Strong interactions - mediated by QCD - have two effects: they i)
distort the spectrum, and ii) let
the atoms decay. As we will see below, strong interactions
may be considered a small perturbation in some cases, and it is then
possible to calculate their effect.
Indeed, as is known since  fifty years\cite{deser}, the energy shift and the lifetime 
of  hadronic  atoms are in general related to the pertinent $T$- matrix element in QCD at
threshold.
 Therefore, measuring the spectra amounts to measure these amplitudes. Classic
 applications of this procedure to determine strong amplitudes are
\bc
\begin{tabular}{lrccl}
Pionic hydrogen &PSI\cite{psi}& $\leftrightarrow$&$T_{\pi N}$ \\
Pionium &DIRAC\cite{tauscher}&$\leftrightarrow$&$T_{\pi\pi}$ \\
Kaonic Hydrogen&DEAR\cite{dear}&$\leftrightarrow$&$T_{\bar KN}$ &\fs
\end{tabular}
\ec
Data on hadronic atoms have therefore the potential to replace low energy ex\-peri\-ments on
\bc
\begin{tabular}{lcc}
$\pi N\rightarrow \pi N$&$\leftrightarrow$&$T_{\pi N}$\\
$\pi\pi\rightarrow \pi\pi$&$\leftrightarrow$&$T_{\pi\pi}$\\
$\bar K N\rightarrow \bar K N$&$\leftrightarrow$&$T_{\bar KN}$
\end{tabular}
\ec
that are difficult (or impossible) to perform. All in all, hadronic atoms
allow one  to confront { high precision, low energy QCD predictions} 
with data. As a now classic example I mention  $\pi\pi$ scattering lengths, where the
theoretical predictions are\cite{bijnenspipi,cgl}
\bea
a_0=0.220\pm 0.005\scs
a_0-a_2=0.265\pm 0.004\scs 
\eea
to be confronted with e.g. data from $K_{e4}$ decay\cite{e865},
\bea
a_0=0.216\pm 0.013\, (\text{stat.})\, \pm0.002\, (\text{syst.})\,\pm0.002\, (\text{theor.})\fs
\eea
Data on $\pi\pi$ scattering from the DIRAC experiment are discussed in
Tauscher's  contribution to
this conference\cite{tauscher}. Furthermore, a high statistics $K_{e4}$
experiment is underway at NA48\cite{na48}. As  Cabibbo has pointed out at
this conference, $K^\pm\rightarrow \pi^\pm\pi^0\pi^0$ decays may provide the possibility to
determine the combination $a_0-a_2$ with high 
precision\cite{cabibbotalk, cabibbo}.

The procedure to confront QCD predictions with data on atomic spectra consists
of two steps:
First, one  relates the spectra to  
QCD scattering amplitudes at threshold\cite{deser}. The precision of this 
 calculation must match the accuracy of the data, which requires in many cases to go beyond
the relation provided in\cite{deser}.  
Second, one calculates QCD amplitudes using effective 
field theories, lattice calculations \ldots, and compares
with what one obtains from step one.

The experimental and theoretical   situation for kaonic atoms 
is summarized in table 1.

\begin{table}[H]
\centering
\vskip.2cm
\renewcommand{\arraystretch}{1.3}
\begin{tabular}{l|r|r}\hline
&{\em { experiment}}&{\em { theory}}\\  \hline
$\bar K\pi$& Letter of Intent\cite{diracloi} &\cite{schweizer,sazdjian}  \\\hline
$\bar Kp$&DEAR \cite{dear} &\cite{ivanovkh,meirusrah}\\\hline
$\bar Kd$&SIDDHARTA\cite{iliescu,jensen} &\cite{ivanovkd}\\\hline
\end{tabular}
\caption{Kaonic atoms: status of theory and experiment.}   
\end{table}
\renewcommand{\arraystretch}{1.0}
\vskip.2cm

The DEAR experiment is presently the only place 
where there is overlap between theory and data in {\em kaonic
  atoms}. Let us hope that the situation changes in the future.

\section{$\bar K \pi $ atoms}
$\bar K \pi $ atoms are interesting, because the hadronic effects  in the spectrum are
 related to $SU(3)\times SU(3)$ chiral perturbation theory (ChPT)  in the meson sector, which works, as far
 as is known today, very well. The modern way to interrelate  the spectrum and
 QCD works as follows. First, one observes that the
momenta of the  constituents  as well as of the decay products are small, of
 the order of 1 MeV or less. Therefore, it is advisable to use a non relativistic
 field theory framework for the calculation\cite{caswell lepage,nreff} - for a
 relativistic approach  see\cite{releff}. In order to verify that
 a perturbative calculation is reasonable, we note that the Coulomb binding energy of
 the ground state is $E_B\simeq 2.9$ keV, whereas the strong shift of the
 energy level is about -9 eV\cite{schweizer} - a tiny effect.
Further, 
the lifetime of the ground state turns out to be about $4\cdot 10^{-15}$ sec.
An  estimate of the number of orbits performed before decaying,
\bea
 \tau E_B\simeq 1.8\cdot 10^4\scs
\eea
 reveals that the atom may be considered as nearly stable. I conclude that
 the calculation is self consistent  - $\bar K\pi$ atoms belong to a
 class of systems where the perturbation of the QED spectrum by the strong
 interaction among the constituents is small. 

Next, we consider the decay channels allowed. The mass differences are
\bea
M_{K^-} + M_{\pi^+} = M_{\bar K^0} + M_{\pi^0} + 0.6 \text{MeV}\scs
\eea
as a result of which possible decay channels are
\bea
A_{K^-\pi^+}\rightarrow \bar K^0\pi^0,\bar K^0+n \gamma,\ldots
\eea
One expands the decay width in powers of the isospin breaking
parameters\footnote{We denote the fine structure constant by $\alpha\simeq 1/137$.}  
$\alpha$ and $ m_d-m_u$, that are counted as quantities of order $\delta$. 
For the ground state, the leading and next-to-leading  terms are due 
to the decay into $\bar K^0\pi^0\,\,\,$: 
\bea
\Gamma_G=\underbrace{a\,\delta^{7/2}+
b\,\delta^{9/2}}_{\bar K^0\pi^0}+\underbrace{
{\cal O}(\delta^5)}_{\bar K^0\pi^0+\text{other channels}}\fs
\eea
 The formula for the decay width of the ground state at next-to-leading order 
 has recently been worked out by
 Julia Schweizer\cite{schweizer},
\bea
\Gamma_G=8\alpha^3\mu_c^2p^*[a_0^-]^2(1+\epsilon) +{\cal O}(\delta^5)\scs
\eea
where
$a_0^-$ is the isospin odd S-wave scattering length in elastic $\pi K$
scattering, 
 $p^*$ denotes the relative 3-momentum of the $\bar K^0 \pi^0$ pair in the final state, and
$\mu_c$ stands for the reduced mass of the charged mesons. Finally, 
the quantity $\epsilon$ is a  correction due to isospin breaking, known at order
$\delta$\cite{schweizer}. Therefore, a measurement of the decay width of the
ground state provides  $a_0^-$,
\bea
\Gamma_G\hspace{.3cm}\rightarrow \hspace{.3cm} { a_0^-}\hspace{.3cm}
\leftrightarrow 
\hspace{.3cm} \text{low energy QCD}\fs
\eea
We note that  $a_0^-$ is the scattering length in pure QCD, purified from
electromagnetic corrections, evaluated at $m_u=m_d$, with $M_K=493.7$ MeV. Using the value of 
$a_0^-$ determined recently in a dispersive analysis\cite{roy steiner}
gives
\bea
\tau_G= (3.7\pm 0.4)\cdot 10^{-15} \text{sec}\fs
\eea
The main open problem here concerns the experimental verification of this
result, and an investigation of whether one may obtain in this manner 
more information on the LECs that occur in the chiral expansion of 
the scattering lengths\cite{bijnenskp}.

 For an exhaustive discussion of the various decay channels and energy
shifts, I refer the interested  reader to the work of 
Julia Schweizer\cite{schweizer}.
I conclude  with the observation that 
the theory of $\pi K$ atoms very well understood. On the other hand,
experiments  are sadly missing.

\section{Kaonic hydrogen}
Here, I discuss properties of kaonic hydrogen, a system investigated in the
last years at DEAR\cite{dear}.
Let us first again discuss  orders of magnitudes. The Coulomb binding energy of the ground
state is about 8.6 keV, the strong shift about .2 keV\cite{dear} -
the perturbation is still small. The width is $\Gamma\simeq 250$
eV\cite{dear}, such that the system performs about 
\bea
\tau\cdot E_B\simeq 35
\eea
 orbits before decaying, considerably less than in the case of the $\pi K$
 atom, but still reasonably many. Note, however, that this number becomes
 $\simeq 10$ for the width found in\cite{meirusrah} from unitarized ChPT 
- which is surprisingly  small.

\subsection{Theory}
Some of the decay channels   of kaonic hydrogen are
\bea
A_{\bar Kp}\rightarrow \pi\Sigma,\Sigma\pi\gamma,\Sigma\pi
e^+e^-,\Sigma\gamma,\ldots
\eea
Note that it cannot decay into an $\bar K^0 n$ pair for kinematic
reasons: in our world, the value of the up and down quark masses are 
such that $M_{K^-}+M_p < M_{\bar K^0}+M_n$. This is in contrast to what happens in
the $\bar K\pi$ atom, where the main decay channel is into the  neutral pair $\bar
K^0 \pi^0$.

\begin{figure}
\centering
\includegraphics[width=10cm]{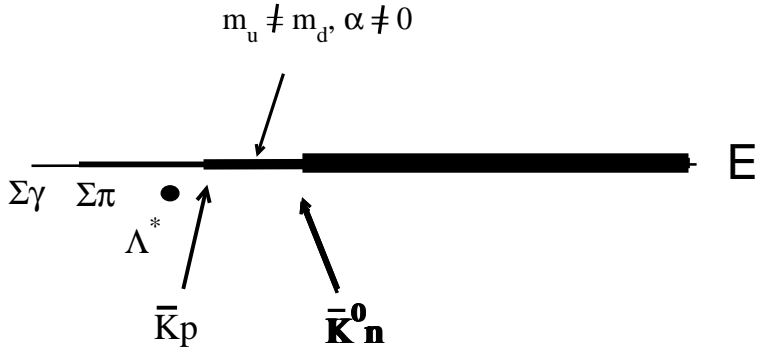}
\caption{The analytic properties of the forward amplitude for $\bar K
  p\rightarrow \bar Kp$ scattering in the presence of isospin breaking
  interactions. Indicated are some of the branch points in the amplitude.
The filled circle denotes the $\Lambda^*(1405)$ pole on the second
  Riemann sheet. The energy axis is not on scale.}\label{fig:analytic} 
\end{figure}

The necessary steps to get the pertinent formula for the
 energy shift and decay width have been performed recently by Mei\ss ner, Rusetsky and 
Raha\cite{meirusrah} in a very nice piece of work in the framework of
 effective field theory, that accounts for a systematic expansion in isospin
 breaking effects. A different approach has been used in\cite{ivanovkh}. In order to
 illustrate the difficulties one is faced with in this system, I display in 
 figure \ref{fig:analytic} the analytic properties of the forward $\bar K p\rightarrow \bar Kp$
 amplitude at $\alpha\neq 0, m_u\neq m_d$. The various branch points and cuts
 have to be taken into account properly in the derivation of the result, and
 this amplitude must then be related to the one in pure QCD, where e.g. the
 branch points at $\bar Kp$ and $\bar K^0 n$ coincide, and where the $\Sigma
 \gamma$ cut is absent. 

The main
observation is the following\cite{meirusrah}: there are large isospin breaking effects in the
final formula, as large as the uncertainty in present DEAR data. Whereas this
observation is not new\cite{dalitz,deloff}, 
the authors of\cite{meirusrah} have shown how to sum up the most
singular pieces, such that the remainder is of
next order in isospin breaking and therefore expected to be small.
 The result for the energy shift and level widths of the 
S-states is similar in structure to the $\bar K\pi$ atom considered above,
 however considerably  more complicated - I
 refer the interested reader to the original article\cite{meirusrah} for the
 explicit formula. The main point is that the shift and width can be
 calculated, once the $I=0,1$ scattering lengths $a_{0,1}$ in $\bar K
 p\rightarrow \bar Kp$ scattering are known in pure QCD, at $m_u=m_d$.
Vice versa, if the shift and width is known, one may determine these
 scattering lengths.

\vskip1cm

\subsection{Comparison with data}
The scattering lengths $a_{0,1}$ have been calculated in\cite{khprediction} - 
see also\cite{khweise,khosetramos} -
by use of  unitarized ChPT. 
 The comparison with the data from the DEAR collaboration is provided in
Ref.~\cite{meirusrah}, to which I refer the reader for details, see in particular
their figure 3, that illustrates the large isospin breaking present in this
system. The theoretical prediction\cite{khprediction}
does not agree with the measurement performed at DEAR - although it must 
be said that the calculation of the scattering lengths 
in\cite{khprediction} does not include an error analysis of the final result.
The reason for this disagreement has not yet been investigated\cite{meirusrah,meissnerkh}.
It is interesting to compare the scattering lengths in\cite{khprediction} 
with ChPT in the standard loop expansion.
The relevant calculation had been performed by 
Kaiser\cite{kaiser}. It turns out that the one loop result for the isospin zero
amplitude is completely off the correct answer, as a result of which the predicted energy shift
in the ground state of kaonic hydrogen has the wrong sign.
This  shows that, due to the nearby resonance $\Lambda^* (1405)$,
 one has to go beyond a pure loop
 expansion. This is what has been done\cite{khweise,khosetramos,khprediction}. However,
 the procedure is not without pitfalls: the authors of e.g. Ref.~\cite{khosetramos}  
have provided scattering lengths  that are in
sharp conflict with the DEAR data. The reason for this failure is explained in\cite{meirusrah}.

Once data on kaonic hydrogen energy shift and width will be available at the eV level, it will be
even more dramatic to compare theoretical prediction with these data - I am
rather curious
to see whether unitarization procedures will pass this test. Needless to say
that it would be comforting to have a precise prediction from theory, including
uncertainties attached, before our experimental colleagues have done their job.
 \footnote{After this manuscript had been submitted for publication in the
  Proceedings, the work of Borasoy et al. has appeared \cite{borasoy},
 which presents a novel theoretical analysis of the strong interaction 
shift and width of kaonic hydrogen in view of the new DEAR measurements~\cite{dear}.}
 Finally, I shortly remind the reader that it would be, in my opinion, 
 a theoretically tremendous effort to derive a precise relation 
between the scattering lengths determined
through the measurement of kaonic hydrogen, and the kaon nucleon sigma 
terms\cite{gasser}.

\section{More complicated systems}
The are more complicated systems than the ones we have considered so far,
e.g., kaonic deuterium. There are plans to investigate this system with
SIDDHARTA, see the contributions by Iliescu and Jensen to this 
conference\cite{iliescu,jensen}.
 The investigation of the relevant spectra can provide 
information on the 
$\bar Kp, \bar Kn$ scattering amplitude at threshold. Of course, one 
needs the corresponding  formula, relating the scattering lengths to 
the spectrum. One may compare this with pionic deuterium, where first 
theoretical investigations using effective field theories are already
available\cite{irgaziev} or underway\cite{rusetsky,rusetskytalk}. The $\bar K d$ system is even more
complicated\cite{ivanovkd}. Whether a theoretically sound analysis in the
framework of effective field theories is possible remains to be seen.

\section{Conclusions}
\begin{enumerate}
\item
{\em Hadronic atoms} are a wonderful tool to measure QCD amplitudes at threshold.
\item
 {\em $\bar K \pi$ atoms} are theoretically well understood\cite{schweizer}\cite{sazdjian}.
 The relevant $ \bar K\pi$ scattering amplitude is now available to two loops 
in ChPT\cite{bijnenskp}, and an analysis
 invoking Roy-Steiner equations has been performed  as well\cite{roy steiner}. 
On the other hand, the precise connection between the vacuum properties of QCD and
 $\bar K\pi$ scattering is still an open question, and experiments on the atom are absent.
\item
 The ground state of {\em kaonic hydrogen} has been investigated in a beautiful 
experiment at DEAR\cite{dear}. Data are available, the S-states state of the atom are
 theoretically understood\cite{meirusrah,ivanovkh}.
\item
On the other hand, the theory of $\bar Kp$ scattering leaves many questions open.
More precise data will reveal whether present techniques are able  to describe 
the complicated situation properly. 
\item
Concerning {\em kaonic deuterium}, experiments are planned\cite{iliescu,jensen}.
 Whether this systems allows for a
theoretically sound analysis in the framework of effective field theory remains to be seen.
\end{enumerate}.

\section*{Acknowledgements}
It is a pleasure to thank the organizers for the invitation to give 
this talk, and
for the very stimulating atmosphere at the conference. 
Furthermore, I thank Ulf-G. Mei\ss ner
and Akaki Rusetsky for illuminating discussions,  Akaki Rusetsky for
 numerical values of scattering lengths evaluated in unitarized chiral
perturbation theory, and Carlo Guaraldo and Akaki Rusetsky
for useful comments concerning the manuscript. This work was 
 supported in part by the Swiss
 National Science Foundation and by RTN, 
BBW-Contract No. 01.0357 
and EC-Contract  HPRN--CT2002--00311 (EURIDICE).

\end{document}